# ASIC-based Implementation of Synchronous Section-Carry Based Carry Lookahead Adders

P. BALASUBRAMANIAN*, N.E. MASTORAKIS[§][¶]
* School of Computer Engineering
Nanyang Technological University
50 Nanyang Avenue
SINGAPORE 639798
Email: balasubramanian@ntu.edu.sg
[§] Department of Computer Science
Military Institutes of University Education
Hellenic Naval Academy
Piraeus 18539, GREECE
Email: mastor@hna.gr
[¶] Department of Industrial Engineering
Technical University of Sofia
Sofia 1000, Boulevard Kliment Ohridski 8
BULGARIA
Email: mastor@tu-sofia.bg

*Abstract:* - The section-carry based carry lookahead adder (SCBCLA) topology was proposed as an improved high-speed alternative to the conventional carry lookahead adder (CCLA) topology in previous works. Self-timed and FPGA-based implementations of SCBCLAs and CCLAs were considered earlier, and it was found that SCBCLAs could help in delay reduction i.e. pave the way for improved speed compared to CCLAs at the expense of some increase in area and/or power parameters. In this work, we consider semi-custom ASIC-based implementations of different variants of SCBCLAs and CCLAs to perform 32-bit dual-operand addition. Based on the simulation results for 32-bit dual-operand addition obtained by targeting a high-end 32/28nm CMOS process, it is found that an optimized SCBCLA architecture reports a 9.8% improvement in figure-of-merit (FOM) compared to an optimized CCLA architecture, where the FOM is defined as the inverse of the product of power, delay, and area. It is generally inferred from the simulations that the SCBCLA architecture could be more beneficial compared to the CCLA architecture in terms of the design metrics whilst benefitting a variety of computer arithmetic operations involving dual-operand and/or multi-operand additions. Also, it is observed that heterogeneous CLA architectures tend to fare well compared to homogeneous CLA architectures, as substantiated by the simulation results.

## 1 Introduction

The carry lookahead adder (CLA) is a logarithmic time adder among the family of high-speed digital adders [1] [2]. The CLA has been implemented in different logic styles such as static CMOS [3], dynamic CMOS [4] [5], all N-transistor logic [6], pass transistor logic [7], adiabatic style energy recovery [8] [9], gate-diffusion input [10] [11] [12], quantum cellular automata (QCA) [13], and using a variety of materials such as gallium arsenide [14], memristor [15] and vertically-stacked nanowire transistors [16] besides the standard Si-based CMOS. Also, diverse design styles were considered for the CLA implementation such as self-timed [17] – [19] and synchronous viz. full-custom and semi-custom ASIC and FPGA [20] – [22]. Further, as a supplement, various low power design strategies such as multiple supply voltages and/or multiple threshold voltages, adiabatic logic, transistor sizing, and transistor reordering have been considered to effect good optimization of the design parameters viz. power, delay, and area.

The design of a CLA is based on the notion that by examining the augend and addend inputs of an adder, it is possible to predict the carry signals in advance thus being able to significantly reduce the linear propagation delay that could be expected in a ripple carry adder (RCA), where carry tends to sequentially propagate from one full adder stage to another. There are two types of CLA architectures, namely homogeneous and heterogeneous. Regular CLAs are also called homogeneous CLAs, and heterogeneous CLAs comprise a mix of CLAs and other carry-propagate adder architectures, for example CLAs and RCAs.

In the rest of this paper, Section 2 discusses the conventional CLA (CCLA) topology and Section 3 discusses the section-carry based CLA (SCBCLA) topology. Section 4 presents the simulation results for 32-bit dual-operand addition corresponding to certain optimized homogeneous and heterogeneous CLA architectures. Finally, the conclusions are made in this section.

## 2 Conventional Carry Lookahead Adder (CCLA)

Assuming $A_i$ and $B_i$ to be the augend and addend inputs of an adder (full adder) stage, and $C_i$ to be its carry input, the (lookahead) carry output viz. $C_{i+1}$ is expressed by (1), and the sum output is expressed by (2). The full adder [23] – [25] is a fundamental arithmetic unit that adds two input bits inclusive of any incoming carry input and produces the sum and carry overflow output. In (1) and (2), $G_i$ and $P_i$ represent generate and propagate signals, where $G_i = A_iB_i$ and $P_i = A_i \oplus B_i$. Product implies logical conjunction, and sum implies logical disjunction in the equations. The symbol $\oplus$ specifies logical exclusivity (i.e. logical XOR). Notice that generate and propagate functions are mutually exclusive – hence, the carry is either generated from an adder stage or the carry simply propagates from input to output. Equations (1) and (2) are inherently in disjoint sum of products (DSOP) or sum of disjoint products form [26] [27]. In such form, any two product terms constituting the Boolean expression would be mutually orthogonal [28], i.e. the logical conjunction of any pair-wise combination of the constituent product terms would result in null.

$$C_{i+1} = G_i + P_iC_i \quad (1)$$

$$Sum_i = P_i \oplus C_i \quad (2)$$

Unwinding the recursion inherent in (1), the carry lookahead signals corresponding to a 3-bit carry lookahead generator are specified by (3) to (5), where $C_0$ represents the carry input to the 3-bit carry lookahead generator, and $C_1$, $C_2$, and $C_3$ represent the corresponding lookahead carry outputs. Note that in a generic m-bit CCLA, a total of 'm' lookahead carry outputs are produced. Equations (3), (4) and (5) show how the lookahead carries are dependent only upon the incoming carry-input to the carry lookahead generator and the corresponding generate and propagate signals, i.e. there is no relation between the intermediate carries. Equations (3), (4) and (5) are also in DSOP form.

$$C_3 = G_2 + P_2C_2 = G_2 + P_2G_1 + P_2P_1G_0 + P_2P_1P_0C_0 \quad (3)$$

$$C_2 = G_1 + P_1C_1 = G_1 + P_1G_0 + P_1P_0C_0 \quad (4)$$

$$C_1 = G_0 + P_0C_0 \quad (5)$$

Fig. 1 portrays the following: Fig. 1a and Fig. 1b depict example homogeneous and heterogeneous 32-bit CLAs; Fig. 1c shows a generic m-bit CCLA (here m = 3), and Fig. 1d shows the internal details of a 3-bit conventional carry lookahead generator which synthesizes (3) to (5). The m-bit CCLA shown in Fig. 1c consists of 3 blocks: (i) propagate-generate logic, which produces propagate and generate signals corresponding to the augend and addend inputs applied, (ii) conventional carry lookahead generator, which accepts the respective propagate and generate signals and the carry input and processes them to produce the lookahead carry outputs including the carry input for the succeeding CCLA according to (1), and (iii) sum logic, which combines the respective propagate and carry signals of the m-bit CCLA according to (2) and processes them to produce the sum outputs of the CCLA.

The 32-bit homogeneous CCLA shown in Fig. 1a is made up of ten 3-bit CCLAs and one 2-bit CCLA in the least significant position. On the other hand, the 32-bit heterogeneous CCLA shown in Fig. 1b comprises ten 3-bit CCLAs and two full adders in the least significant position. From Fig. 1a it should be clear that the 32-bit homogeneous CCLA is composed using smaller size CCLAs alone, while the 32-bit heterogeneous CCLA is constructed using smaller size CCLAs and an RCA. The use of a 2-bit RCA in the least significant position helps to somewhat reduce the critical path delay, area, and power dissipation parameters of the 32-bit heterogeneous CCLA in comparison with the 32-bit homogeneous CCLA. This will be substantiated by the simulation results given in Section 4.

For larger CCLAs, besides considering an RCA that is composed using 1-bit full adders in the least significant position of the heterogeneous CCLA architecture, an RCA implemented using 2-bit (i.e. dual-bit) full adders [29] may also be considered for

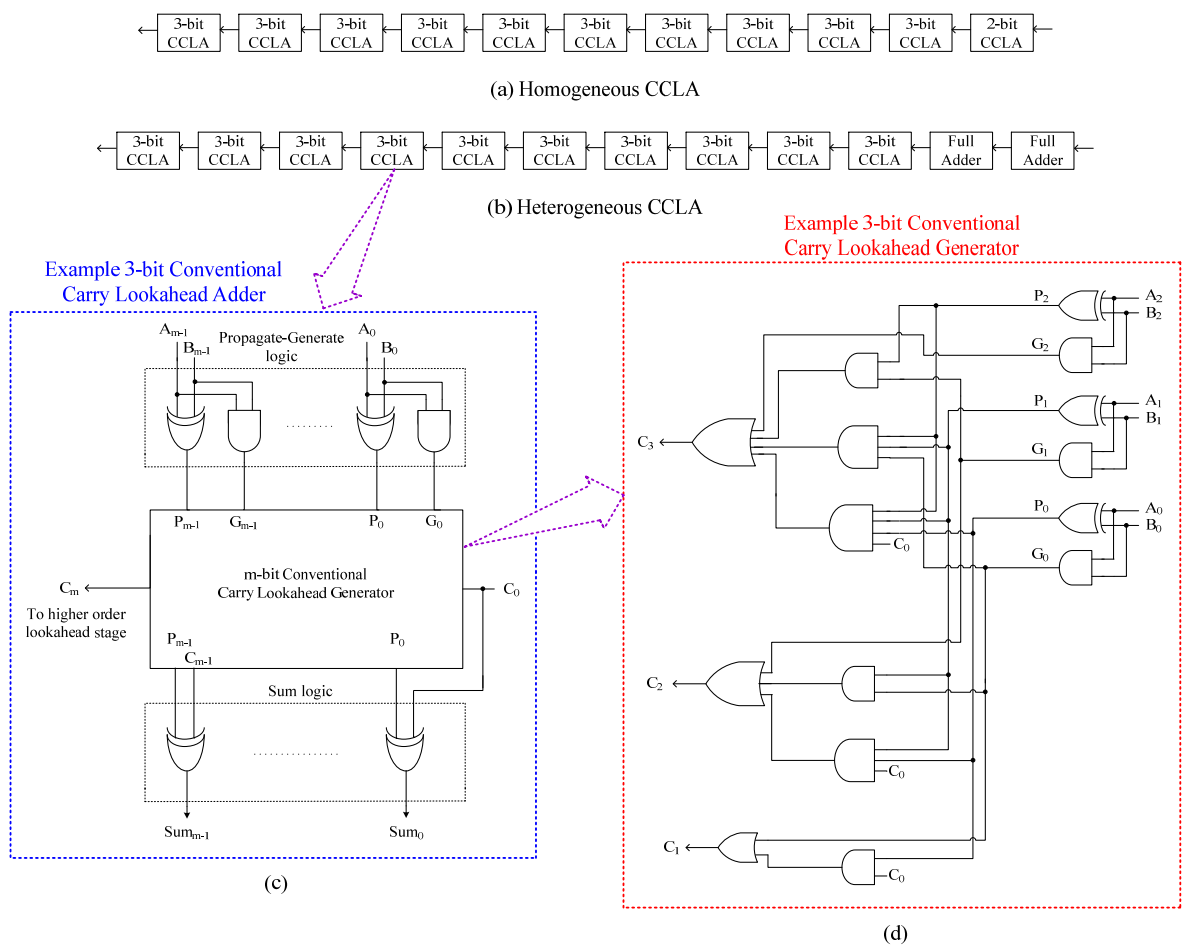

Fig. 1 (a) Example 32-bit homogeneous CCLA; (b) Example 32-bit heterogeneous CCLA; (c) Example m-bit CCLA (where m = 3); (d) Example 3-bit conventional carry lookahead generator the heterogeneous CCLA architecture which might help to better optimize the design parameters. Nevertheless, the optimal use of dual-bit full adders in the RCA vis-à-vis using single-bit full adders in the RCA should be verified through pre-layout simulations in order to better understand the design tradeoffs between employing single-bit full adders and dual-bit full adders based RCA in the heterogeneous CCLA architecture.

## 3 Section-Carry Based Carry Lookahead Adder (SCBCLA)

The SCBCLA is a variant of the CCLA in that only one lookahead carry output, which could serve as the carry input for the successive SCBCLA stage is alone determined. The computation of sum outputs of a SCBCLA are not dependent upon the internal lookahead carry outputs, rather they are produced on the basis of the sequential rippling of the carry signal from one full adder stage to another within the SCBCLA (i.e. sub-SCBCLA). Therefore the SCBCLA produces a lookahead carry output on the basis of the carry lookahead adder architecture, while producing the sum outputs on the basis of the ripple carry adder architecture. As mentioned earlier for the case of CCLAs, pure SCBCLAs are also called homogeneous SCBCLAs, and heterogeneous SCBCLAs feature SCBCLAs and another adder architecture (here, RCA). Further, heterogeneous SCBCLAs are likely to feature optimized design metrics compared to homogeneous SCBCLAs.

Some example 32-bit heterogeneous SCBCLAs are shown in Fig. 2. Similar input and output labels have been adopted in Fig. 2e and 2f as that of Fig. 1c and 1d for ease of comparison. Notice that while the propagate-generate logic of the SCBCLA shown

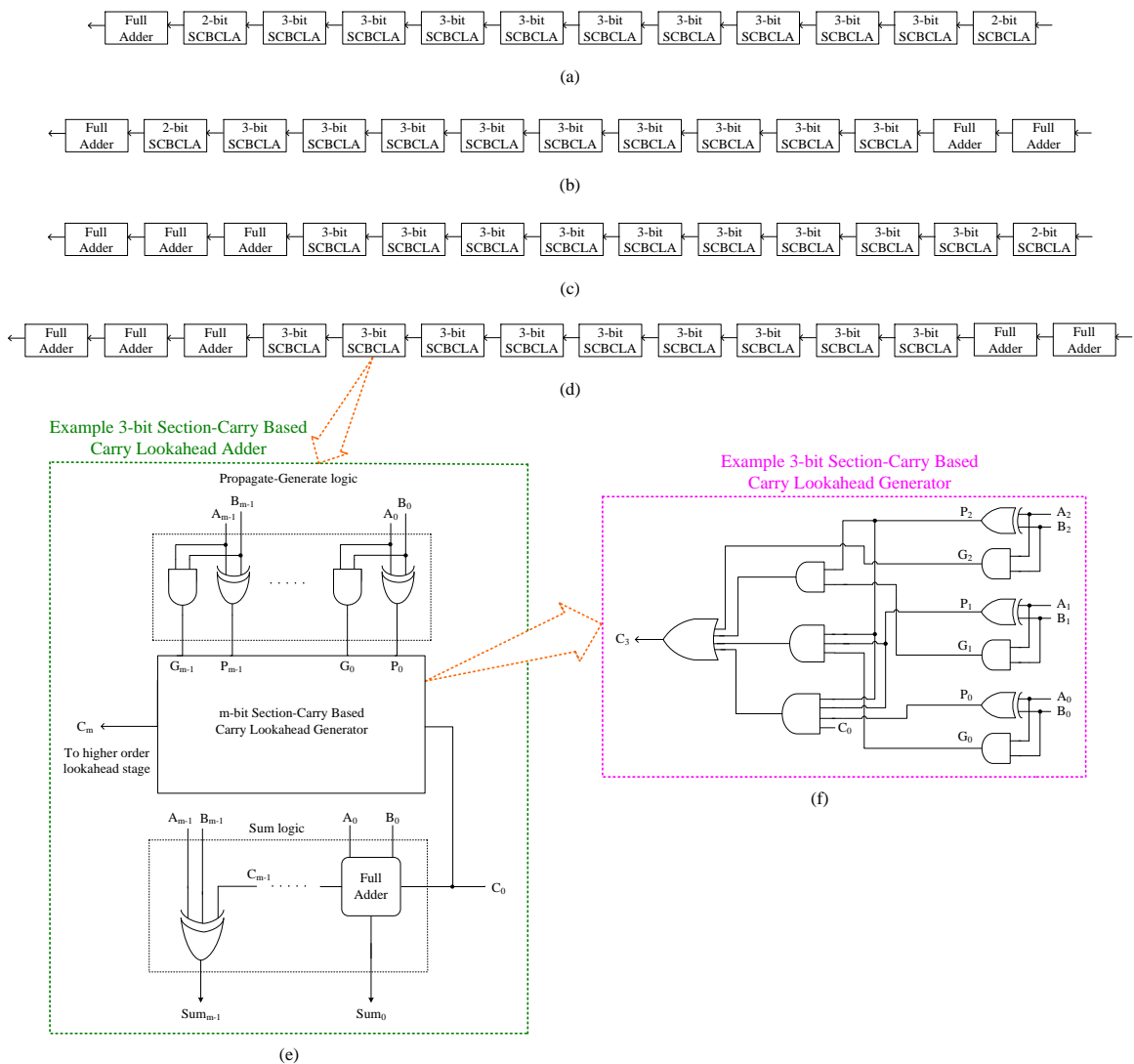

Fig. 2 (a), (b), (c), (d) Example 32-bit heterogeneous SCBCLAs; (e) Example m-bit SCBCLA (where m = 3); (f) Example 3-bit section-carry based carry lookahead generator in Fig. 2e is identical to that of the CCLA shown in Fig. 1c, the sum logic of the SCBCLA shown in Fig. 2e is different from the sum logic of the CCLA shown in Fig. 1c. This is because the sum outputs of the CCLA are computed according to (2), while the sum outputs of the SCBCLA are produced based on the rippling of the carry signal from one full adder stage to another within a sub-SCBCLA. Also, note that the most significant sum output of a sub-SCBCLA is produced through a 3-input XOR gate instead of the full adder, as shown in Fig. 2e, since there is no requirement to keep track of the carry overflow bit.

Fig. 2f shows the lookahead carry output logic of a 3-bit SCBCLA, which synthesizes (3). Comparing this with the 3-bit conventional carry lookahead generator shown in Fig. 1d, it can be seen that both the conventional carry lookahead generator and the section-carry based carry lookahead generator logic are identical, and that in the case of the latter, the intermediate lookahead carry outputs alone are not present, i.e. the logic required to produce the intermediate lookahead carry outputs viz. $C_2$ and $C_1$ are absent in Fig. 2f unlike in Fig. 1d. In fact, while a generic m-bit CCLA will feature 'm' lookahead carry outputs, any m-bit SCBCLA will feature only one lookahead carry output.

Due to the absence of intermediate lookahead carry output(s) in the SCBCLA contrary to the CCLA, the size of the former is expected to be less than the size of the latter. The 2-bit CCLA occupies an area of $32.02\mu m^2$ while the 2-bit SCBCLA occupies an area of only $19.31\mu m^2$ (i.e. 40% reduced area). The 3-bit CCLA occupies an area of $53.12\mu m^2$ while the 3-bit SCBCLA occupies a reduced area of just $29.48\mu m^2$ (i.e. 45% reduced area). These area estimates correspond to a 32/28nm CMOS digital cell library [30]. The area estimates confirm the compactness of the SCBCLA against the CCLA. When the lookahead size is increased from 2-bits to 3-bits, the CCLA suffers from an area increase of 66%, and the SCBCLA experiences an enhanced area expenditure of about 53%.

## 4 Simulation Results and Conclusions

A cell-based semi-custom implementation of various homogeneous and heterogeneous 32-bit CCLAs and SCBCLAs, which were shown in Fig. 1a, 1b and Fig. 2a, 2b, 2c and 2d was considered, by utilizing the elements of the 32/28nm CMOS digital cell library [30]. The power, delay and area results obtained are shown in Table 1.

Table 1. Average power dissipation, critical path delay, and Silicon area occupancy of different 32-bit CLAs

| Type of CLA and Design Legend | Power (µW) | Delay (ns) | Area (µm²) |
|---|---|---|---|
| Homogeneous CCLA (Fig. 1a) – Design 1 | 38.11 | 2.22 | 563.18 |
| Heterogeneous CCLA (Fig. 1b) – Design 2 | 37.60 | 2.18 | 540.82 |
| Heterogeneous SCBCLA (Fig. 2a) – Design 3 | 42.99 | 2.20 | 483.64 |
| Heterogeneous SCBCLA (Fig. 2b) – Design 4 | 41.92 | 2.16 | 462.03 |
| Heterogeneous SCBCLA (Fig. 2c) – Design 5 | 42.23 | 2.27 | 462.03 |
| Heterogeneous SCBCLA (Fig. 2d) – Design 6 | 41.16 | 2.23 | 440.43 |

The design metrics estimated correspond to a typical case PVT specification with recommended supply voltage of 1.05V and operating junction temperature of 25ºC. For estimating average power dissipation, more than 1000 random input vectors were identically supplied to the different CLAs at time intervals of 5ns (200MHz) through a test bench. The .vcd files generated through the functional simulations were subsequently used for accurate average power estimation using Synopsys PrimeTime by invoking the time-based power analysis mode. The maximum propagation delay (i.e. critical path delay) and Silicon area were also estimated with suitable wire loads included automatically whilst performing the simulations. Minimum-sized discrete and complex gates (full adders) of the cell library [30] were chosen uniformly for the different CLA designs and the CLA outputs were assigned with fanout-of-4 drive strength. This paves the way for a straightforward comparison of the design metrics of different CLAs subsequent to physical synthesis.

To comprehensively comment on the design parameters of different CLAs, a figure-of-merit (FOM) is defined as the inverse of the product of power, delay, and area (i.e., $PDAP^{-1}$) [31] – [36]. Since minimization of power, delay, and area is desirable, a lower PDAP value and thus a higher FOM value can be considered to be an indicator of an optimized design. The calculated FOM values of various 32-bit CLAs which are scaled up by a factor of $10^6$ are mentioned in Fig 3 using the design legends specified in Table 1.

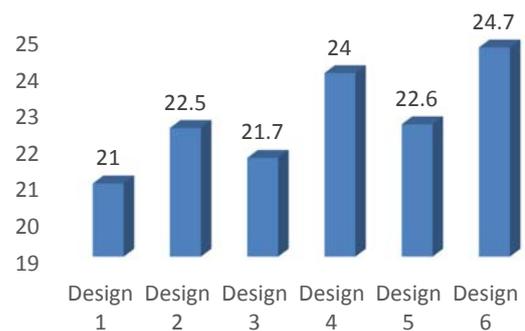

Fig 3. FOM (i.e. $PDAP^{-1}$) of various 32-bit CLAs viz. CCLAs and SCBCLAs

From Fig 3, it can be seen that the homogeneous CCLA has the least FOM amongst the various designs considered. In comparison with it, the heterogeneous CCLA reports a 7% increase in the FOM. The calculated FOM values shown in Fig 3 duly convey that heterogeneous CLA architectures are preferable from the design viewpoint compared to homogeneous CLA architectures. Moreover, the FOM values portrayed in Fig 3 clearly indicate that the heterogeneous SCBCLA designated as Fig. 2d has the best FOM amongst all its counterparts. Compared to the FOM values of homogeneous and heterogeneous CCLAs, the proposed heterogeneous SCBCLA (shown as Fig. 2d) reports corresponding increases in FOM by 17.6% and 9.8% respectively. In terms of peak power dissipation though, the homogeneous 32-bit CCLA featured the least value of 6.34mW, with the heterogeneous SCBCLA corresponding to Fig 2a dissipating the maximum value of 11.99mW. Nevertheless, average power dissipation is more important for assessing the low power attribute of a digital design.

From an ASIC-based design perspective, the two important conclusions drawn from this research are: (i) the SCBCLA architecture is more optimized compared to the CCLA architecture, and (ii) the heterogeneous CLA architecture is preferable compared to the homogeneous CLA architecture to achieve good optimization of the design metrics with respect to both CCLA and SCBCLA. These two conclusions were found to hold well in the case of previous works [18] [19] which dealt with ASIC-based self-timed implementations of CLAs. Hence, it is opined that the two conclusions derived are likely to hold good for semi-custom ASIC-based implementation of higher-order synchronous CLAs as well.